\documentclass[twocolumn]{aastex63}
\shorttitle{3D dissipative pulsar magnetospheres}
\shortauthors{Cao \& Yang}
\begin{document}
\title{Three-dimensional dissipative pulsar magnetospheres with Aristotelian electrodynamics}
\author{Gang Cao}
\affiliation{Department of  Mathematics, Yunnan University of Finance and Economics, Kunming 650221, Yunnan, P. R. China, gcao@ynufe.edu.cn}
\author{Xiongbang Yang}
\affiliation{Department of Astronomy, Yunnan University, Kunming 650091, Yunnan, P. R. China}

\begin{abstract}
A good compromise between the resistive model and the PIC model is Aristotelian electrodynamics, which can include the back-reaction of the radiative photons onto particle motion and allow for a local dissipation where the force-free condition is violated. We study the dissipative pulsar magnetosphere with Aristotelian electrodynamics where particle acceleration is fully balanced by radiation. The expression for the current density is defined by introducing a pair multiplicity. The 3D structure of the pulsar magnetosphere is then presented by solving the time-dependent Maxwell equations using a pseudo-spectral algorithm. It is found that the dissipative magnetosphere approaches the force-free solution and the dissipative region is  more restricted to the current sheet outside the light-cylinder (LC) as the  pair multiplicity increases.  The spatial extension of the dissipative region  is self-consistently controlled by the pair multiplicity. Our simulations show the high magnetospheric dissipation outside the LC  for the low pair multiplicity.
\end{abstract}


\section{Introduction}
Pulsars are rapidly rotating and highly magnetized neutron stars, which lose their rotational energy through the electromagnetic radiation. They can produce the broadband electromagnetic
spectrum throughout the entire electromagnetic spectrum from the radio to $\gamma$-ray bands. The radiation from these objects is thought to originate from the high-energy particles accelerated by unscreened electric fields. The accelerated high-energy particles flow along the open field lines and produce the synchrotron, curvature, and inverse Compton  radiation.
The Fermi Gamma-Ray Space Telescope launched in 2008 has opened a new perspective  on the pulsar physics with more than 100 detected gamma-ray pulsars \citep{abd10,abd13}.
The Fermi observations provide the valuable information about pulsar light curves, phase-averaged spectra and phase resolved spectra. However, it is still unclear where the particles in the magnetosphere are accelerated and how their radiation is produced. This requires us to have a deeper understanding of the precise pulsar magnetosphere. In fact, the particle accelerations in the magnetosphere are associated with the magnetosphere electrodynamics, which requires a self-consistent calculation of
the  Maxwell equations by including particle dynamics and radiation to model the global pulsar magnetosphere. The modeling of pulsar magnetospheres have achieved significant
progress over the last decade.

The  vacuum dipole field is the first solution of the global pulsar magnetosphere, which is widely used as the pulsar background field  and the radiation modeling in the early stage of pulsar study.
The advantage of the vacuum dipole model is that an exact analytical solution is available and is known as the \citet{deu55} solution.
The standard gap models including the polar cap \citep[e.g.,][]{rud75,dau82}, the slot-gap (SG) \citep[e.g.,][]{dyk03,dyk04,mus04}, and the outer-gap (OG) \citep[e.g.,][]{che86,zha97,che00} models are based on this field structure. Such gap models have achieved great successes on the pulsar radiation modeling \citep[e.g.,][]{wat09,rom10,pet19}.
The vacuum model does not take into account the effects of the current on the magnetosphere structure. In fact, the magnetospheric current will significantly change the field structures outside the light cylinder(LC). Therefore, the vacuum solution is not a real pulsar model.

The pulsar magnetosphere is  filled with plasma created by pair cascades \citep{gol69}. When the density of plasma is enough high, any accelerating electric fields is shorted out so that the condition ${\bf E}\cdot {\bf B}=0$ holds everywhere. This is referred to as force-free electrodynamic and corresponds to the zeroth-order approximation of the plasma-filled magnetosphere. The force-free solution has recently been achievable with the development of the numerical method.
The first numerical solution to the force-free magnetosphere for an aligned rotator is found by \citet{con99}. The CKF solution consists of the open and closed field lines inside the LC and an equatorial current sheet outside the LC. The axisymmetric force-free solution was further explored by the time-dependent simulations by several groups \citep[e.g.,][]{kom06,mck06,tim06,yu11,par12,pet16,cao16a,eti16,car18}. These time-dependent simulations also confirmed the closed-open CKF solution with an equatorial current sheet outside the LC.
\citet{spi06} firstly presented the three-dimensional (3D) structures of the force-free pulsar magnetosphere for the oblique rotator. The 3D force-free solution was further studied based on different numerical algorithm by several groups \citep{kal09,pet12,tch13,cao16b}. They reproduced the closed-open CKF solution with an equatorial current sheet extending to several LC.
In fact, the force-free solutions are  dissipationless by definition, meaning that they do not allow  any particle acceleration and production of radiation in the magnetosphere. Therefore, the force-free solution is also not a true pulsar model.
\begin{figure*}
\center
\begin{tabular}{cccccccccccccc}
\includegraphics[width=5.4 cm,height=5.5 cm]{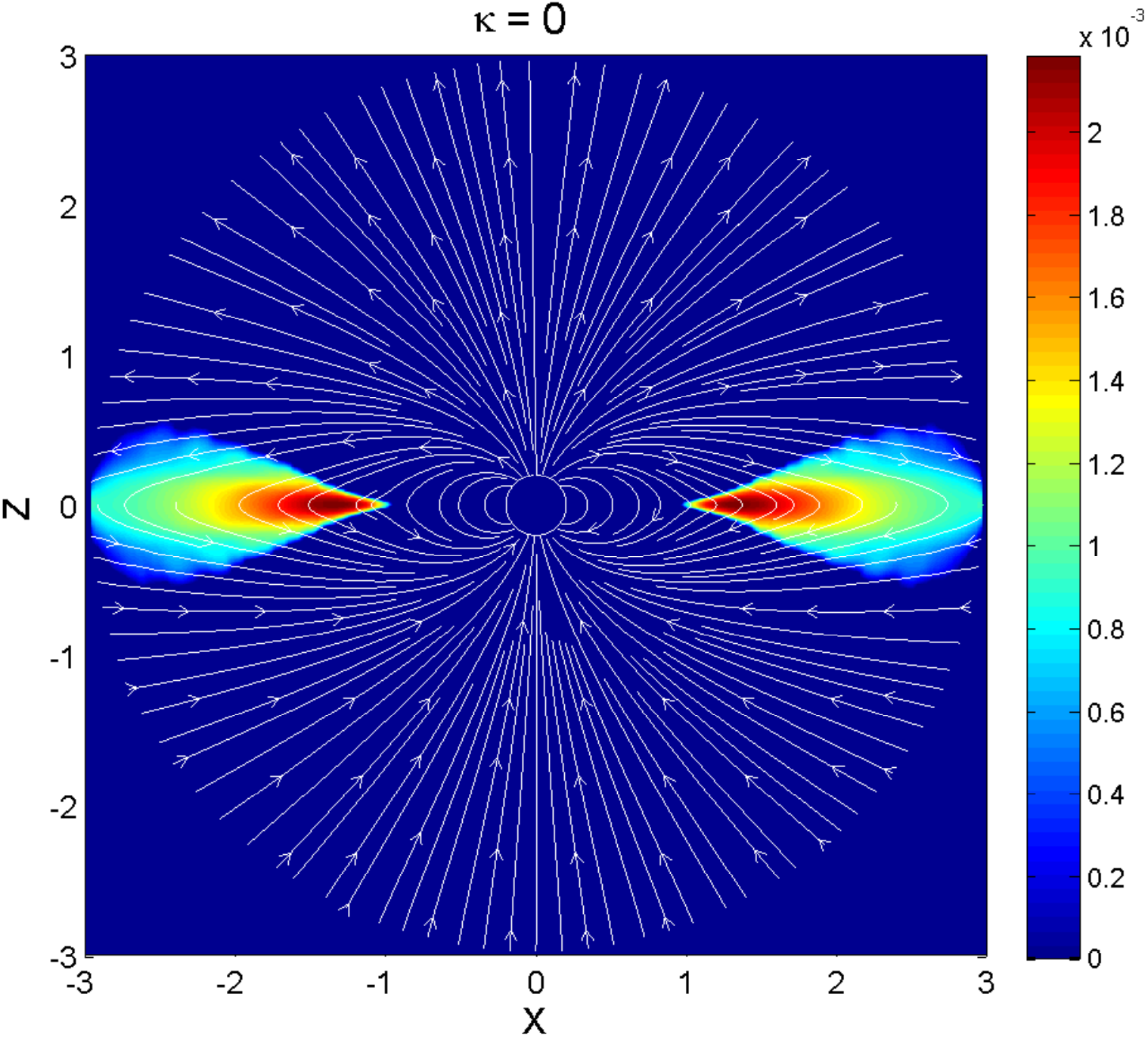} \qquad
\includegraphics[width=5.4 cm,height=5.5 cm]{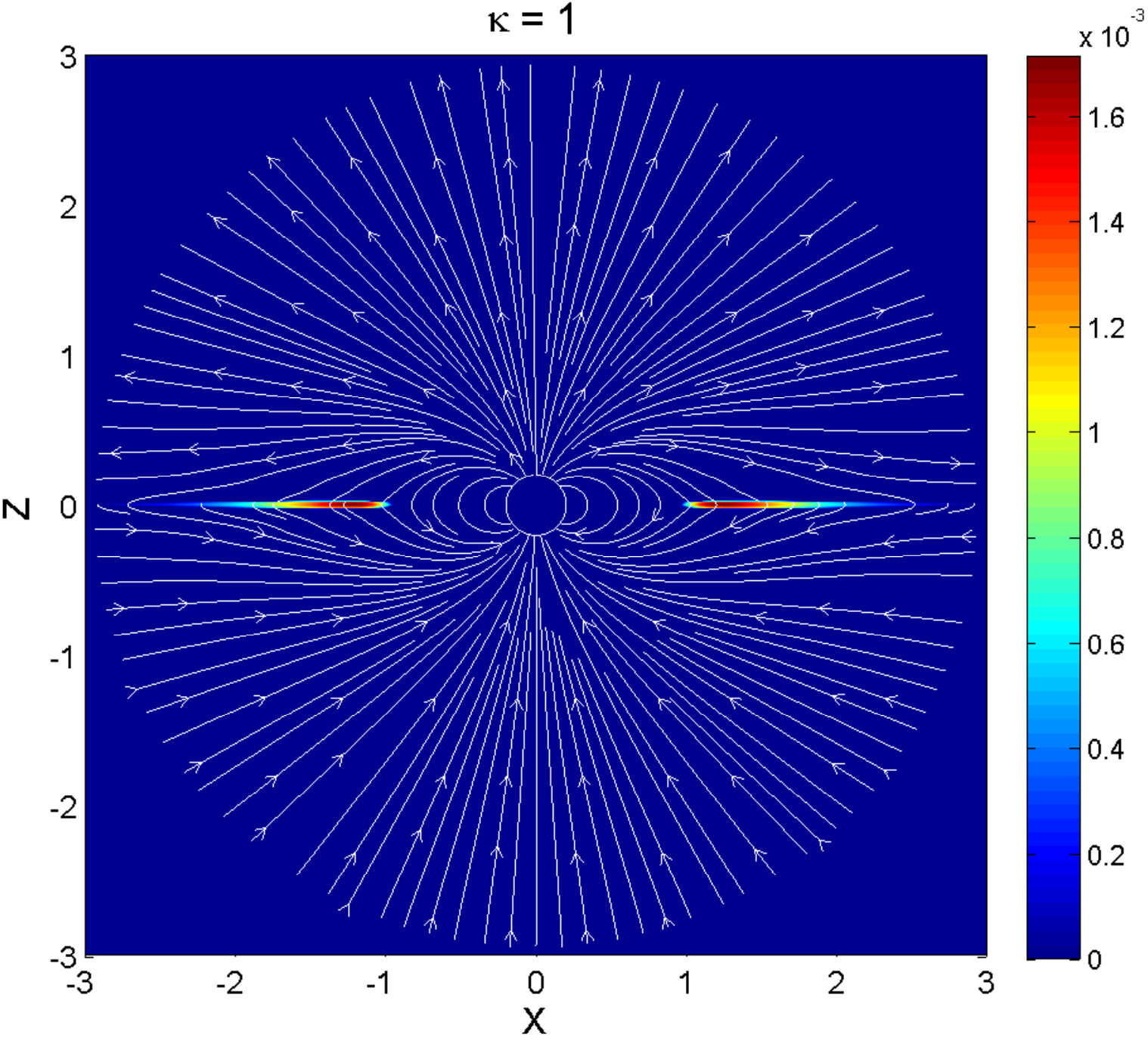} \qquad
\includegraphics[width=5.4 cm,height=5.5 cm]{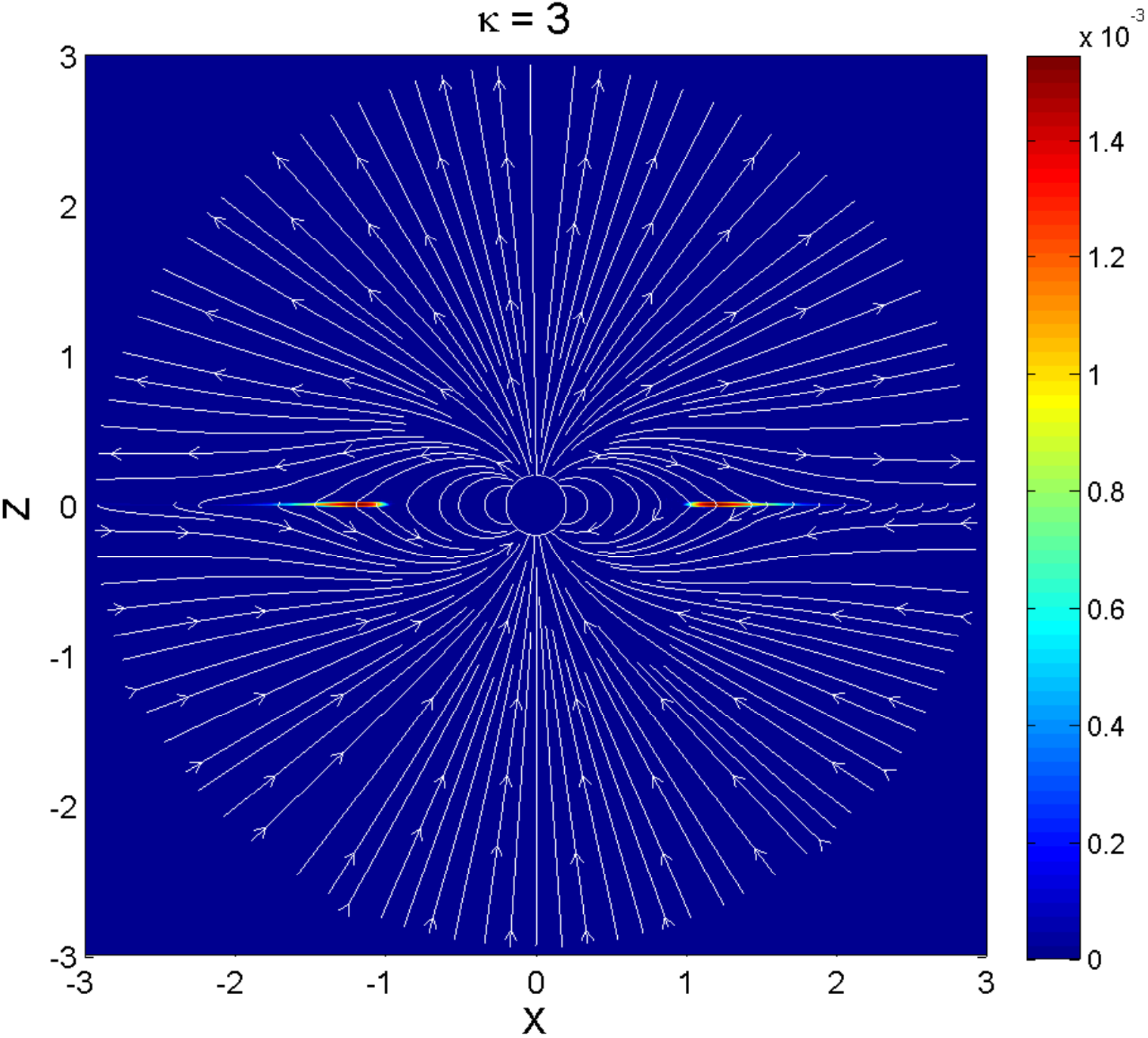} \qquad
\end{tabular}
\caption{Distribution of the magnetic field lines and the accelerating electric field $E_0$ in the x-z plane for an aligned dissipative rotator with the pair multiplicity $\kappa=\{0,1,3\}$.\\}
\end{figure*}

The realistic pulsar magnetosphere should have some dissipation regions to allow for the particle acceleration. The dissipative pulsar magnetosphere with plasma resistivity have been explored by \citet{li12}, \citet{kal12a} and \citet{cao16b}. The resistive model usually requires a macroscopic conductivity parameter to define the current density and control the accelerating electric field. The introduction of a finite conductivity can span the magnetospheric solutions  from the vacuum to force-free field. The resistive pulsar magnetospheres are also used to predict the pulsar light curves and energy spectrum \citep{kal14,bra15,kal17,cao19,yan19}. These studies revealed that the particle acceleration and high-energy emission is produced near  the current sheets.
Recently, particle-in-cell (PIC) methods with the self-consistent feedback between particles and fields are used to model the pulsar magnetosphere \citep{phi14,che14,bel15,cer15,phi15,kal18,bra18} and predict the pulsar light curves \citep{cer16,phi18,kal18}. However, the PIC simulation can not adjust the particle energy to  the realistic $\gamma$-ray emitting particle energy.\\

Another way to introduce the dissipation in the magnetosphere is to use the radiation reaction limit, which is also called  Aristotelian electrodynamics. The radiation reaction limit have been used to model the pulsar magnetosphere \citep{gru13,con16,pet20}. \citet{con16} explored the radiative magnetospheres based on radiative magnetospheres for an oblique rotator.
However, they did not include the  current density along the magnetic field. This description can not reflet the real Aristotelian electrodynamics. Recently,  \citet{pet20} presented the structure of radiative pulsar magnetospheres by including the full current description but only for the axisymmetric rotator. Also, an alternative derivation of the particle velocities in the equatorial current sheet is presented  which turns out to be equivalent to the Aristotelian prescription \citep{con20}.
In this paper, we study the dissipative pulsar magnetosphere with  Aristotelian electrodynamics for the oblique rotator. We present the 3D structure of the dissipative pulsar magnetospheres by  a pseudo-spectral method.
The paper is organized as follows: In section 2, we describe the model of Aristotelian electrodynamics. In section 3, we present the results from our simulation. Finally, a brief discussion and conclusions are given in section 4.

\section{Aristotelian electrodynamics}
The time-dependent Maxwell equations are
\begin{eqnarray}
{\partial {\bf B}\over \partial t}=-{\bf \nabla} \times {\bf E}\;,\\
{\partial  {\bf E}\over \partial t}={\bf \nabla} \times {\bf B}-{\bf J}\;,\\
\nabla\cdot{\bf B}=0\;,\\
\nabla\cdot{\bf E}=\rho_{\rm e}\;,
\end{eqnarray}
where ${\bf J}$ is the current density and $\rho_{\rm e}$ is the charge density.  The time-dependent Maxwell equations can be solved by implementing a prescription for the current density ${\bf{J}}$.

Pulsar magnetospheres are loaded  with  electron/positron pairs. These particles can be accelerated to relativistic energy by unscreened accelerating field and radiate photons in all wavelengths.
These photons have a back-reaction onto the particle motion and make the particles brake in a direction opposite to their motion. It is expected that the particle acceleration and radiation can reach a stationary balance in the magnetosphere, which is called  radiation reaction limit or Aristotelian electrodynamics. In Aristotelian electrodynamics, the radiation reactions have a different way to
electrons and positrons. The velocity for two types of particles can be described by the local electromagnetic field \citep{fin86,gru13}
\begin{eqnarray}
{\bf v_{\pm}}=  {{\bf E} \times {\bf B}\pm(B_0{\bf {B}}+E_0{\bf {E}}) \over B^2+E^2_{0}},
\label{Eq6}
\end{eqnarray}
where the two signs correspond to positrons and  electrons, they  move at the speed of light in the magnetosphere. The quantities $B_0$ and $E_0$ is defined by the
Lorentz invariants
\begin{eqnarray}
B^2_{0}-E^2_{0}={\bf B}^2-{\bf E}^2,\,\,\, E_{0}B_{0}={\bf E}\cdot {\bf B},\,\,\,E_{0}\geq0,
\end{eqnarray}
where $B_0$ and $E_0$ are the magnetic and electric field in the frame in which ${\bf E}$ and ${\bf B}$ are parallel.
The quantity $E_0$ is  the effective accelerating electric, which is zero when ${\bf E}\cdot {\bf B}=0$.

\begin{figure}
\epsscale{0.98}
\plotone{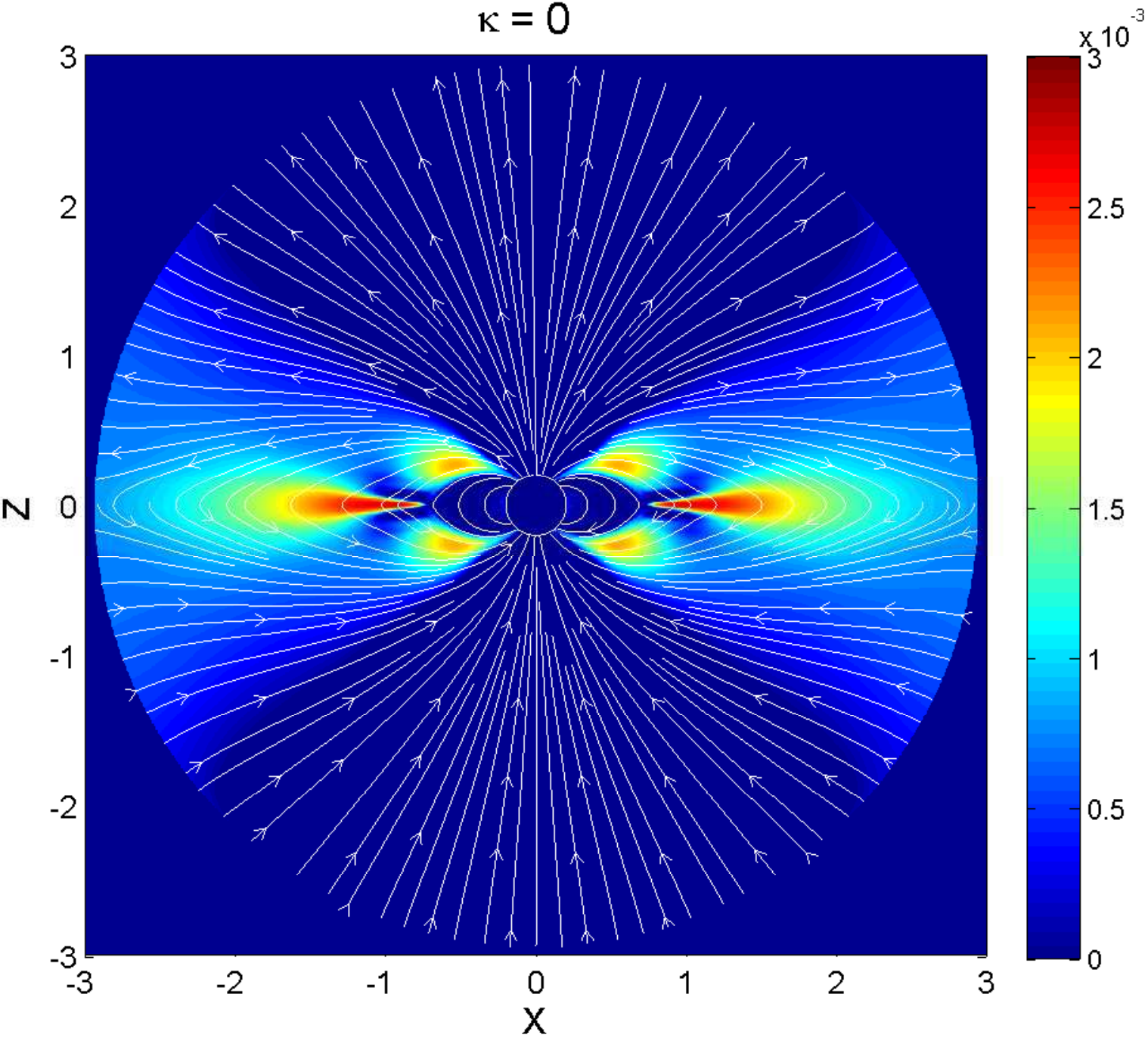}
\caption{Same as figure 1, but where the AE formulation is applied in the whole magnetosphere. }
\end{figure}

\begin{figure}
\center
\begin{tabular}{cccccc}
\includegraphics[width=7.2cm,height=6.5cm]{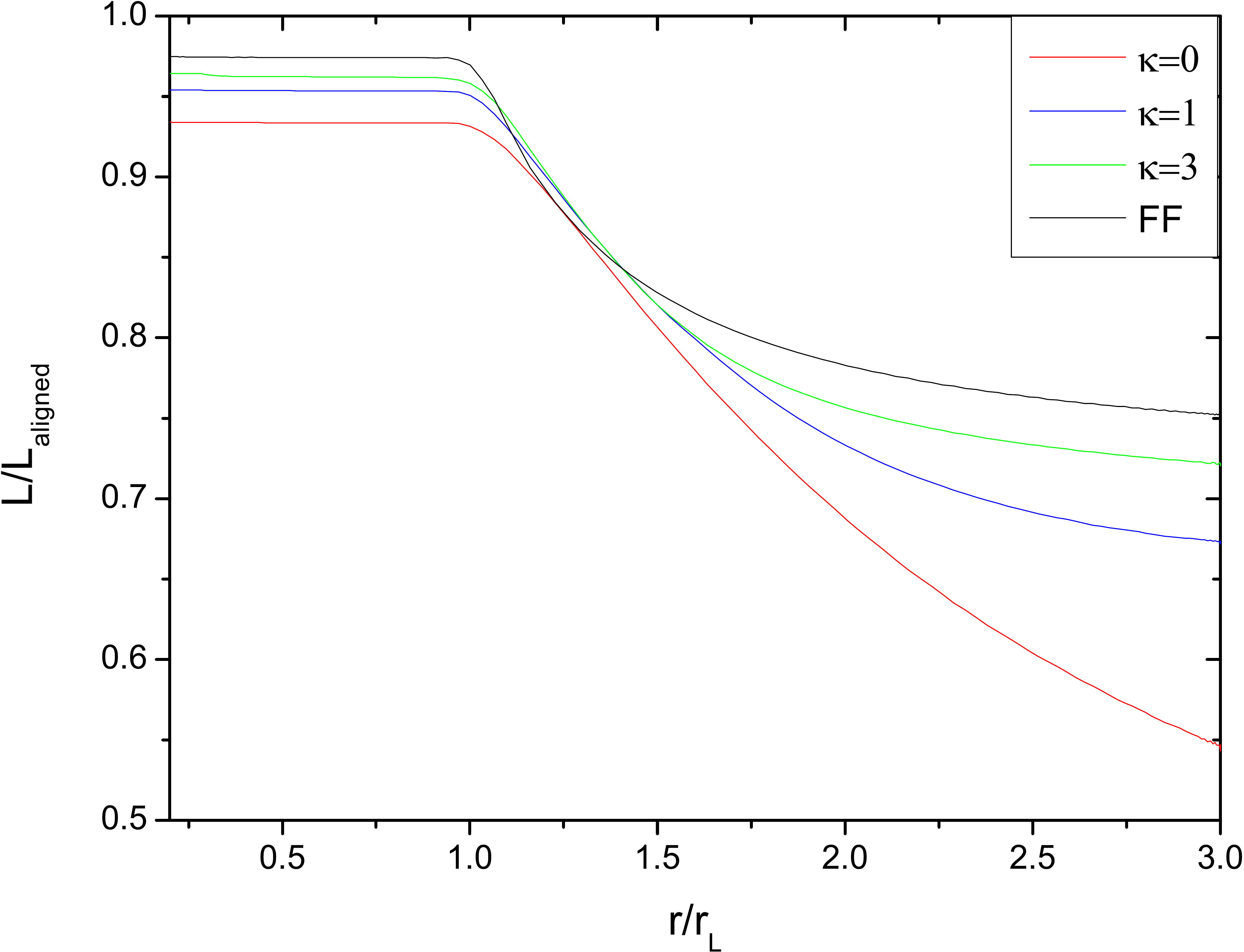}
\end{tabular}
\caption{Normalized Poynting flux $L/L_{\rm aligned}$ as a function of radius $r$ for an  aligned dissipative rotator with different pair multiplicities $\kappa$.}
\end{figure}

The current density can be derived by equation (5) as
\begin{eqnarray}
{\bf J }=   {\rho_e {\bf E} \times {\bf B} + \rho_0 (B_0{\bf {B}}+E_0{\bf {E}}) \over B^2+E^2_{0}},
\label{Eq6}
\end{eqnarray}
where $\rho_0=\rho_{+}+\rho_{-}$ is the total charge density. Equation (7) is not a  form of Ohm¡¯s law, we need to give a description for the total charge density $\rho_0$.
We  define the current density by introducing  the pair multiplicity $\kappa$ as
\begin{eqnarray}
{\bf J }=  \rho_e  \frac {{\bf E} \times {\bf B}}{B^2+E^2_{0}}+ (1+\kappa)\left|\rho_e\right| \frac{ (B_0{\bf {B}}+E_0{\bf {E}}) }{ B^2+E^2_{0}}.
\label{Eq6}
\end{eqnarray}
When the pair multiplicity $\kappa=0$,  the current density is exactly consistent with the one given by \citet{gru13}.
In fact, there is no unique prescription for the current density since we  only know the  total charge density $\rho_0 \geq \left|\rho_e\right|$.
Therefore, we guess the total charge density to be $(1+\kappa)\left|\rho_e\right|$ by introducing the  pair multiplicity $\kappa$.
It is noted that the current density that we used in equation (8) is different from that of \citet{pet20}.  We also obtain a similar result for the aligned rotator by using the current density given by  \citet{pet20}.

\begin{figure*}
\center
\begin{tabular}{cccccccccccccc}
\includegraphics[width=5.4 cm,height=5.5 cm]{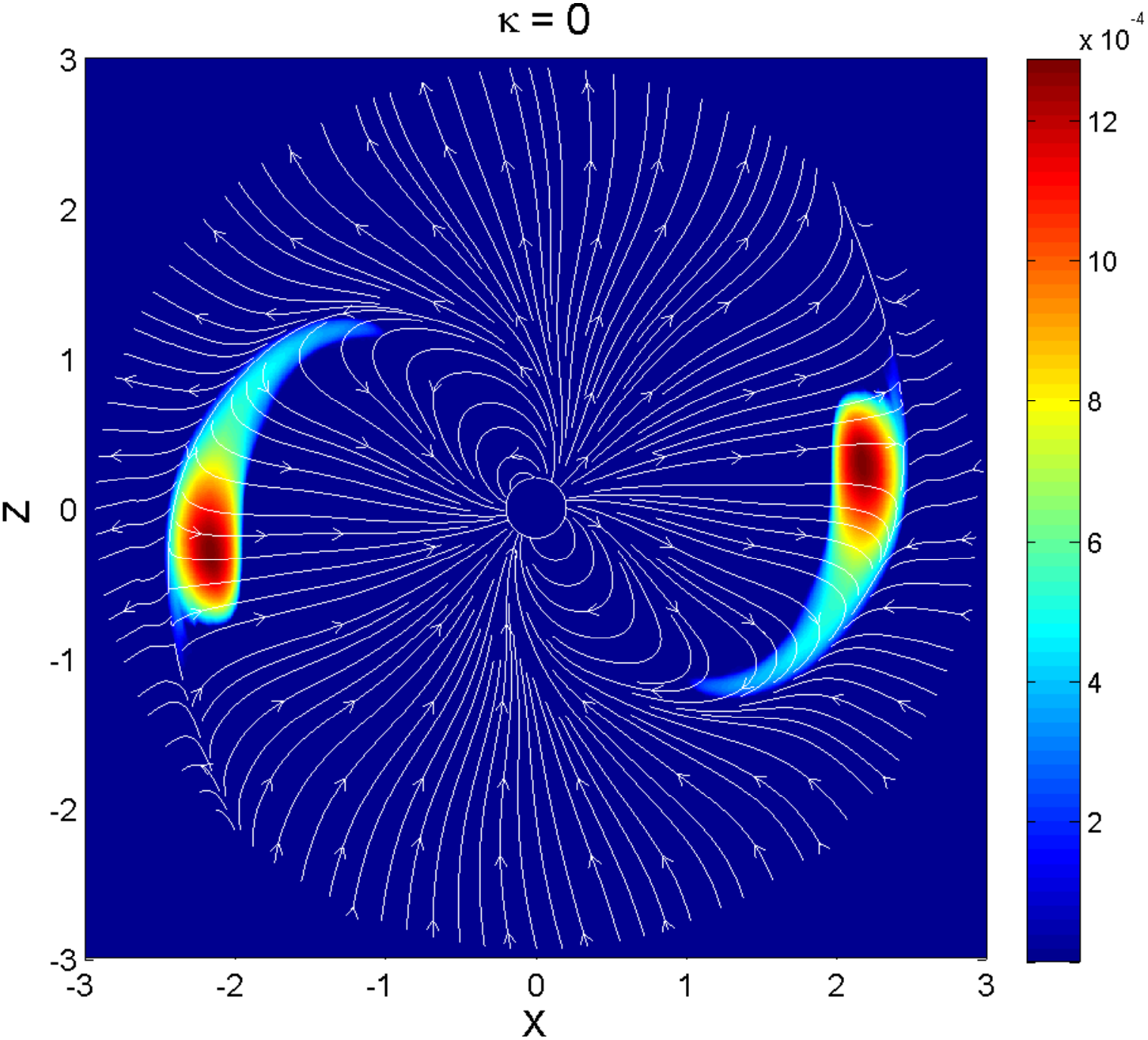} \qquad
\includegraphics[width=5.4 cm,height=5.5 cm]{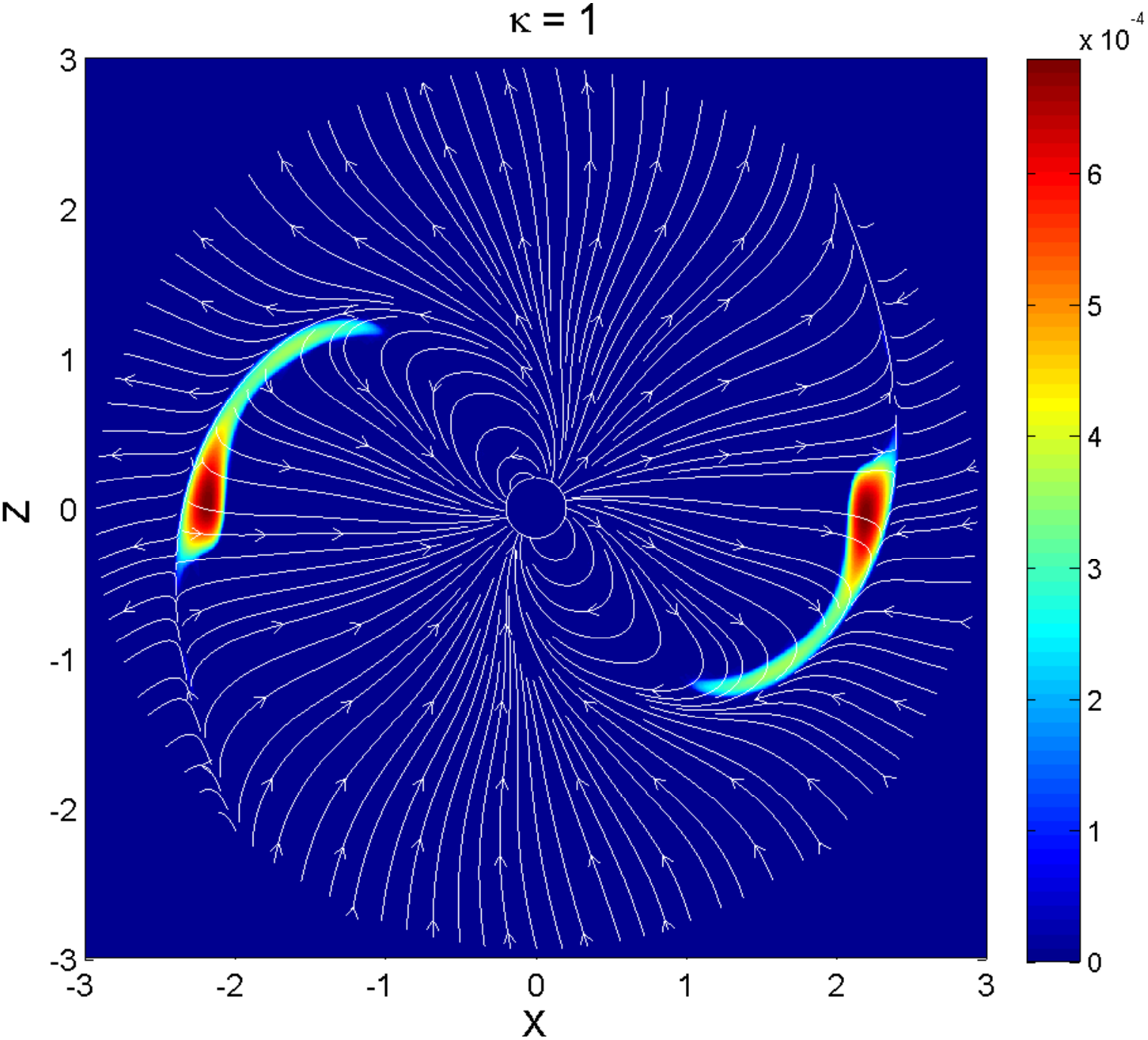} \qquad
\includegraphics[width=5.4 cm,height=5.5 cm]{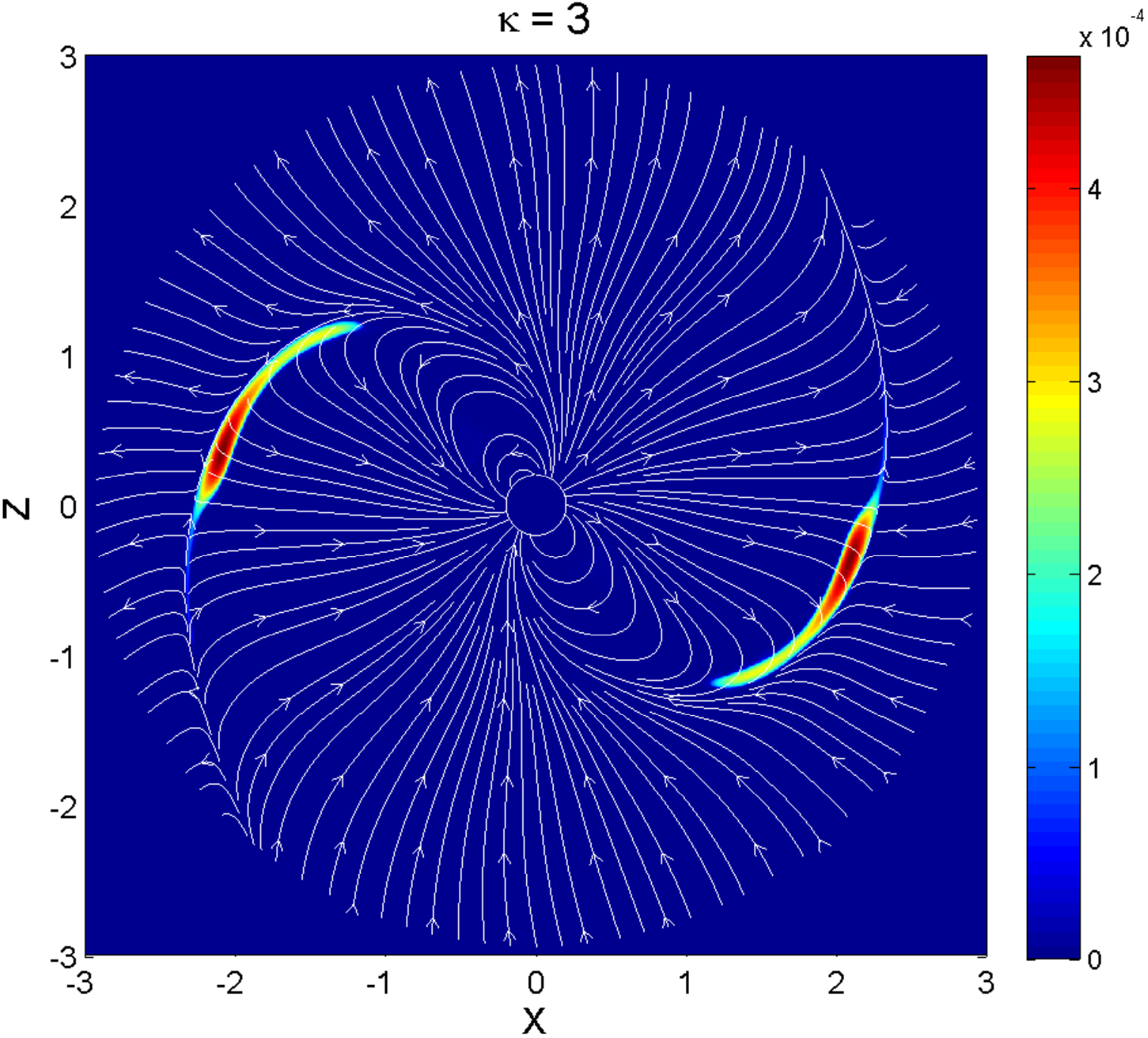} \qquad
\end{tabular}
\caption{Distribution of the magnetic field lines and the accelerating electric field $E_0$ in the x-z plane for a $\alpha=60^{\circ}$ dissipative rotator with the pair multiplicity $\kappa=\{0,1,3\}$.\\  }
\end{figure*}

A current sheet appears in the magnetosphere for the force-free electrodynamics, which is captured by enforcing  the condition $E = B$ in the regions where $E>B$. However,
this condition does not naturally come from the force-free equations and is artificially imposed to avoid the drift current to become superluminal. There is no reason to require that
$E \leq B$ in the current sheet. Moreover, the force-free approximation can not allow for any dissipation and thus preclude the possibility of particle acceleration and the pulsed emission  in the magnetosphere. This impossibility comes from the force-free condition ${\bf E}\cdot {\bf B}=0$. We should leave the force-free description to accommodate the acceleration of particle and the production of radiation  in the magnetosphere. In fact, the Aristotelian electrodynamics can allow for a local dissipative region where $E > B$ .
To explore the dissipation in the region where $E > B$, we enforce the force-free condition in the region where $E \leq B$. This makes the magnetic field lines close and forms a death zone within the LC.

\section{result}
We use a pseudo-spectral method to solve the time-dependent Maxwell equations with our new prescription for the current density.
A set of the spectral collocation points are used to discretize the electromagnetic field in spherical coordinates ($r$,\,$\theta$,\,$\phi$).
A Chebyshev expansion is used in the radial coordinate $r$ and a vector spherical harmonic expansion is used in the angle coordinates ($\theta,\phi$).
The divergencelessness of magnetic field is analytically enforced by a projection method.  A spectral filter in all directions is used  to reduce aliasing errors and Gibbs oscillation. We use a third-order Adam-Bashforth integrator to advance the solutions at each time step. For a detailed description about the pseudo-spectral algorithm, see \citet{cao16a} and \citet{cao16b}.
We initialize the  magnetic field  to an oblique vacuum dipole with magnetic inclination angle $\alpha=\{0^\circ, 30^\circ, 60^\circ, 90^\circ\}$. The inner boundary condition  is enforced at the stellar surface with a rotating electric field ${\bf {E}} = -( {\bf \Omega } \times {\bf r} ) \times {\bf B}/c$. A non-reflecting boundary condition is implemented to avoid the inward reflection from the outer boundary. The computational domain is set to  $r\in (0.2 - 3)$ $r_{\rm L}$. A good accuracy can be obtained with a typical resolution of $N_r \times N_{\theta} \times N_{\phi}=128 \times 32 \times 64$ . We performed several simulations with the dissipative magnetospheres for the pair multiplicity $\kappa=\{0,1,3\}$.
We evolve the system  for several rotational periods so that a final steady solution can be reached.

We show the structure of magnetic field lines and the distribution of the accelerating electric field $E_0$ in the x-z plane for  an aligned rotator  with the pair multiplicity $\kappa=\{0,1,3\}$ in figure 1. Inside the LC, the field lines are similar  and insensitive to the $\kappa$ value, since there is no dissipation as $E < B$ within the LC.
When $\kappa=0$, the field lines close well beyond the LC and are more similar to an aligned dipole field. Outside the LC, we observe  an extended $E_0$ distribution where $E>B$ along the equator.
As  the pair multiplicity $\kappa$ increases, the features of the force-free solution start appearing, the field lines open gradually and become more radial beyond the LC, an equatorial current starts forms outside the LC. We also observe that the $E_0$ regions decreases with increasing pair multiplicity and are more restricted to only near the equatorial current sheet outside the LC for  a moderate pair multiplicity $\kappa\gtrsim1$.
For comparison, we also show the magnetic field lines and the $E_0$ distributions for the pair multiplicity $\kappa=0$ by implementing the  AE formulation everywhere in figure 2.
The field structure is similar to the ``device" found by \citet{gru14} with  a co-rotation zone, a force-free zone and a dissipative zone. We see an extended $E_0$ distribution near the co-rotation zone within the LC. A strong $E_0$ region also exists along the equator outside the LC. This result is  different from the previously obtained one in figure 1.
It is noted that the separatrix return currents are absent in Aristotelian electrodynamics when $\kappa=0$ , which is very different from the force-free one. A similar result is also found
by \cite{con16}.
We show the normalized Poynting flux $L/L_{\rm aligned}$ as a function of radius $r$ for an aligned rotator with different pair multiplicities in figure 3. We see that the Poynting flux increases with increasing  pair multiplicity and approaches the force-free solution for high $\kappa$ value inside the LC. We observe a significant dissipation outside the LC.
About $40 \% $ of the Poynting flux is dissipated  for the pair multiplicity $\kappa=0$ within $3 \, r_L$. The dissipative rate decreases with increasing $\kappa$ value and decreases to $\sim20 \% $ for the force-free solution. The dissipated energy is converted into  particle acceleration and radiation in the dissipative region. Our results are similar to those of the PIC simulation with an increase of particle injection rate \citep{kal18}.

The global  magnetospheric structures for the oblique rotator  are very similar to the aligned one. We show the structure of magnetic field lines and the distribution of the accelerating electric field $E_0$ in the x-z plane for  a $60^\circ$ rotator  with the pair multiplicity $\kappa=\{0,1,3\}$ in figure 4.  As the pair multiplicity $\kappa$ increases,
the field structure tends to the force-free solution with an equatorial current sheet outside the LC. We observer a dissipative region where $E>B$ outside the LC.
The spatial extension of the dissipative region decreases with increasing pair multiplicity and the $E_0$ region  is more confined to the equatorial current sheet outside the LC as the  pair multiplicity $\kappa$ increases. In fact, the $E_0$ distribution for the high $\kappa$ solution is qualitatively similar to the FIDO one \citep[see, e.g.,][]{kal14,cao19}.
We also compare the field structures for  $\kappa=0$  with Fig. 1 of Contopoulos 2016 for $\alpha=0^\circ$ and $\alpha=60^\circ$ respectively. We find that the field structures are  qualitatively very similar to those of \citet{con16}.
For comparison, we also show the magnetic field lines and the $E_0$ distributions for a $60^\circ$ rotator with the pair multiplicity $\kappa=0$ by implementing the  AE formulation everywhere in figure 5. The magnetospheric structure is very similar to the aligned one with a force-free zone bounded by a radiation zone. We observe a strong $E_0$ distribution inside the LC, which is very different from those in the SG and OG models. A strong $E_0$ region with $E>B$  also appears outside the LC.
We show the distributions of magnetic field lines and the accelerating electric field $E_0$ in the x-z plane for  a $30^\circ$ rotator  with the pair multiplicity $\kappa=3$
in figure 6. We see that the field structure is very close to the force-free one and the $E_0$ region is restricted to only near the current sheet outside the LC for this high $\kappa$ value.
We also show the normalized Poynting flux $L/L_{\rm aligned}$ as a function of radius $r$ for a  $90^\circ$ rotator with different pair multiplicities in figure 7. We see that the Poynting flux increases with increasing $\kappa$ values  and approaches the force-free solution for the high $\kappa$ value. Our simulation shows a more than $1\%$ dissipation rate outside the LC for a $90^\circ$ dissipative rotator. A similar dissipation rate is also found  by the PIC simulation for the aligned and perpendicular rotator \citep{phi15}.
In fact, the spectral numerical methods present an unphysical dissipation beyond the LC due to discontinuity in the current sheet. A higher resolution is necessary to catch the discontinuity in the current sheet and reduce the unphysical dissipation.
\begin{figure}
\epsscale{0.98}
\plotone{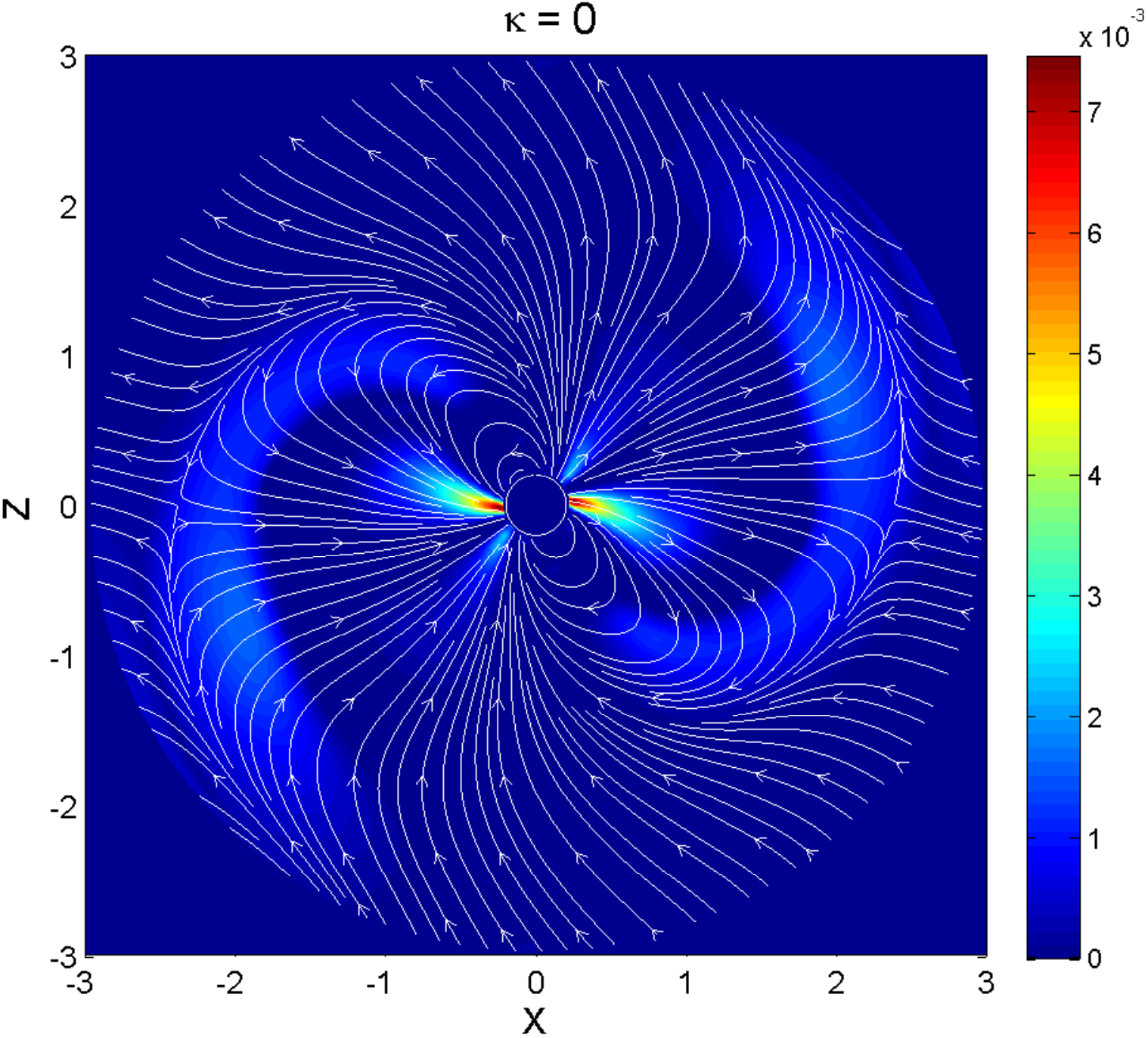}
\caption{Same as figure 4, but where the AE formulation is applied in the whole magnetosphere.}
\end{figure}

\begin{figure}
\epsscale{1.}
\plotone{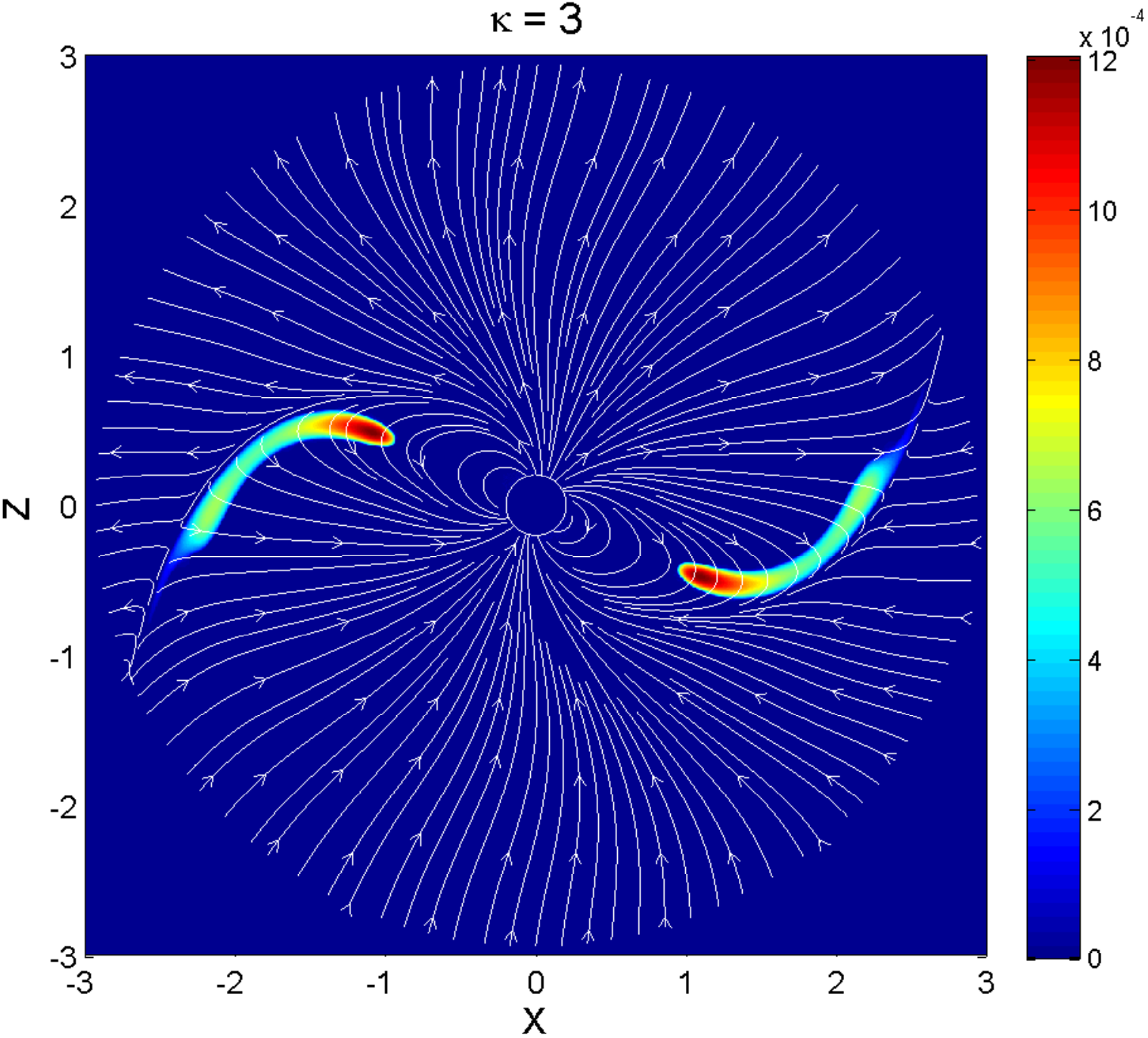}
\caption{Same as figure 4, but for a  $30^\circ$ rotator with the pair multiplicities $\kappa=3$.}
\end{figure}

\begin{figure}
\center
\begin{tabular}{cccccc}
\includegraphics[width=8.cm,height=7.cm]{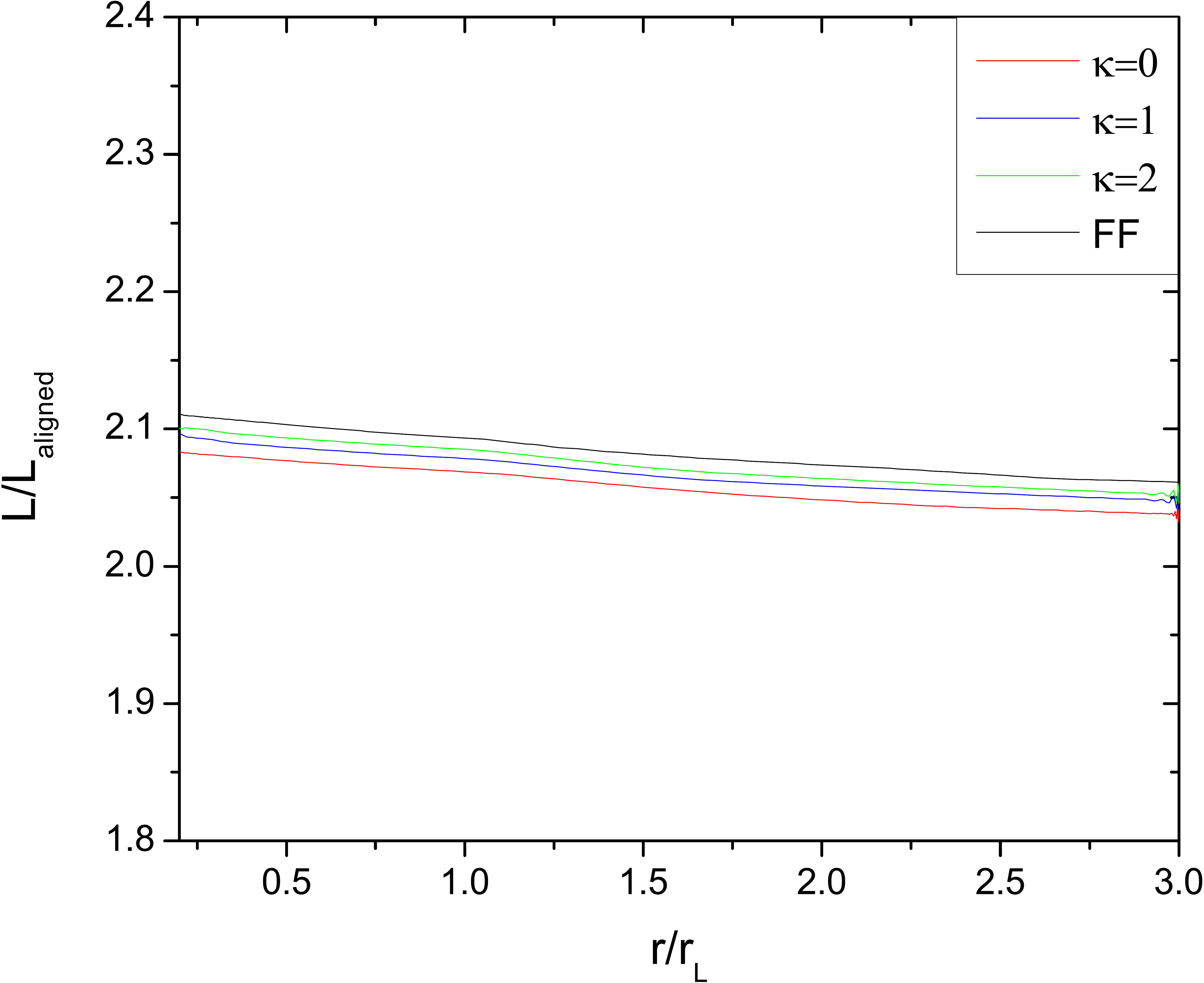}
\end{tabular}
\caption{Normalized Poynting flux $L/L_{\rm aligned}$ as a function of radius $r$ for a  $90^\circ$ rotator with different pair multiplicities $\kappa$. }
\end{figure}

\section{Discussion and Conclusions}
In this paper, we study the dissipative pulsar magnetosphere with Aristotelian electrodynamics where particle acceleration is fully balanced by radiation. We define  the current density as a form of the  Ohm's law by introducing a pair multiplicity. We then present the 3D structure of pulsar magnetosphere  by solving the time-dependent Maxwell equations using a pseudo-spectral algorithm. We find that the dissipative magnetosphere tends to the force-free solution and the dissipative region is more confined to only near the equatorial current sheet for a moderate pair multiplicity $\kappa\gtrsim1$. The spatial extension of the dissipative region  is self-consistently controlled by the pair multiplicity. In fact, our results are in qualitative agreement with those of the recent PIC simulation.

The force-free model can not allow for any dissipation in the magnetosphere. Realistic pulsar magnetosphere should allow for a local dissipation in the magnetosphere to accommodate the acceleration of particle and the production of radiation. The resistive model can produce the magnetic dissipation by relaxing the force-free condition. However, there is no back-reaction of emission onto particle dynamics in the resistive model. The recent PIC method attempts to model pulsar magnetospheres by including the self-consistent feedback between particle motions and electromagnetic fields. The present PIC codes cannot catch all the physics from macroscopic scales to microscopic scales.
A good compromise between the resistive model and  PIC model is Aristotelian electrodynamics, which includes the back-reaction of the emitting photons onto particle motion and allows for the dissipation where the force-free conditions is violated. In fact, the accelerating electric field distribution from our simulation is very similar to that of the PIC simulation with increasing  particle injection in the magnetosphere \citep{kal18}. Our simulation shows a higher magnetic dissipation for the low pair multiplicity and a lower magnetic dissipation for the high pair multiplicity in the aligned rotator,  which is similar to those found in the PIC simulation with particle injection only from the stellar surface and  abundant particle injection in the whole magnetosphere \citep{phi14,che14,cer16}. In fact, it is too restrictive and ad hoc for imposing a charge density by a pair multiplicity. We will solve a full set of the AE equations by including the charge continuity equation with the pairs injection based on the spectral method in the near future.
The pulsar $\gamma$-ray data from  Fermi observation can be used to  constrain the dissipative regions and radiation mechanisms in the magnetosphere. In the next step,
we use the presented dissipative solution to model pulsar $\gamma$-ray light curves and energy spectrum. We expect this study to enhance our understanding of the physical mechanisms behind the high-energy emission in pulsar magnetospheres.

\acknowledgments
We thank the  referee Ioannis Contopoulos for valuable comments and suggestions.
We would like to thank  J\'{e}r$\hat{\rm o}$me P\'{e}tri for some useful discussions.  We acknowledge the financial support from the National Natural Science Foundation of China 11573060 and 11661161010, the National Science Foundation of China 11673060,  the National Science Foundation of China 11871418.


\end{document}